\documentclass{article}

\usepackage{arxiv}
\usepackage{float} 
\usepackage{amsmath}
\usepackage[utf8]{inputenc} % allow utf-8 input
\usepackage[T1]{fontenc}    % use 8-bit T1 fonts
\usepackage{hyperref}       % hyperlinks
\usepackage{url}            % simple URL typesetting
\usepackage{booktabs}       % professional-quality tables
\usepackage{amsfonts}       % blackboard math symbols
\usepackage{nicefrac}       % compact symbols for 1/2, etc.
\usepackage{microtype}      % microtypography
\usepackage{lipsum}
\usepackage{graphicx}
\graphicspath{ {./images/} }

\usepackage{tikz}
\usetikzlibrary{quantikz2}
\usepackage{braket}
\usepackage{subcaption}

\title{Quantum Attention for Vision Transformers in High Energy Physics}

\author{
  Alessandro Tesi \\
  University of Pavia \\
  Pavia, 27100, Italy \\
  \texttt{alessandro.tesi02@universitadipavia.it} \\
  \And
  Gopal Ramesh Dahale \\
  Indian Institute of Technology Bhilai \\
  Bhilai, 491001, Chhattisgarh, India \\
  \texttt{gopald@iitbhilai.ac.in} \\
  \And
  Sergei Gleyzer \\
  Department of Physics \& Astronomy \\
  University of Alabama \\
  Tuscaloosa, AL 35487, USA \\
  \texttt{sgleyzer@ua.edu} \\
  \And
  Kyoungchul Kong \\
  Department of Physics \& Astronomy \\
  University of Kansas \\
  Lawrence, KS 66045, USA \\
  \texttt{kkong@ku.edu} \\
  \And
  Tom Magorsch \\
  Physik-Department \\
  Technische Universität München \\
  Garching, 85748, Germany \\
  \texttt{tom.magorsch@tum.de} \\
  \And
  Konstantin T. Matchev \\
  Institute for Fundamental Theory, Physics Department \\
  University of Florida \\
  Gainesville, FL 32611, USA \\
  \texttt{matchev@ufl.edu} \\
  \And
  Katia Matcheva \\
  Institute for Fundamental Theory, Physics Department \\
  University of Florida \\
  Gainesville, FL 32611, USA \\
  \texttt{matcheva@ufl.edu} \\
}

\begin{document}
\maketitle
\begin{abstract}
We present a novel hybrid quantum-classical vision transformer architecture incorporating quantum orthogonal neural networks (QONNs) to enhance performance and computational efficiency in high-energy physics applications. Building on advancements in quantum vision transformers, our approach addresses limitations of prior models by leveraging the inherent advantages of QONNs, including stability and efficient parameterization in high-dimensional spaces. We evaluate the proposed architecture using multi-detector jet images from CMS Open Data, focusing on the task of distinguishing quark-initiated from gluon-initiated jets. The results indicate that embedding quantum orthogonal transformations within the attention mechanism can provide robust performance while offering promising scalability for machine learning challenges associated with the upcoming High Luminosity Large Hadron Collider. This work highlights the potential of quantum-enhanced models to address the computational demands of next-generation particle physics experiments.
\end{abstract}

% keywords can be removed

\section{Introduction}

The anticipated launch of the High Luminosity Large Hadron Collider (HL-LHC) \cite{HL_LHC_Project} by CERN at the end of this decade is expected to generate an unprecedented volume of data, necessitating advanced computational frameworks and strategies to handle, process, and analyze this immense dataset efficiently. Classical computing resources, while effective, face significant limitations in scaling to the data and computational demands projected by such high-dimensional tasks. Addressing this challenge, quantum machine learning (QML) \cite{QML1, QML2} has emerged as a promising solution.

Quantum vision transformers (QViTs) \cite{guo2024quantumlinearalgebraneed, qvitulu, Comajoan_Cara_2024, Cherrat_2024} have recently been proposed as hybrid architectures that integrate quantum circuits within classical vision transformer (ViT) \cite{dosovitskiy2021imageworth16x16words} frameworks to reduce time complexity and improve performance in machine learning tasks involving high-dimensional data. Traditional ViTs employ self-attention mechanisms \cite{vaswani2023attentionneed} and multi-layer perceptrons (MLPs) \cite{MLP} to learn from image data, which has shown promising results in computer vision tasks across various domains. To advance these models further, researchers have explored replacing the classical linear projection layers in the self-attention mechanisms with ansatz quantum circuits (VQCs), a strategy designed to harness quantum computation for increased efficiency in parameter optimization and feature extraction.

Our work builds on this quantum-classical hybrid framework by utilizing quantum orthogonal neural networks (QONNs) \cite{kerenidis2022classicalquantumalgorithmsorthogonal, orthonn1}. This modification offers a fundamental advancement by enabling inherently orthogonal transformations that provide stability and improved gradient properties in the high-dimensional data spaces characteristic of high-energy physics. The orthogonality of QONNs allows for more efficient learning and enhances model robustness, especially beneficial in contexts where data complexity and noise pose significant challenges, such as jet classification in particle physics.

To demonstrate the efficacy of our QViT model based on QONNs, we apply it to the problem of distinguishing between quark-initiated and gluon-initiated jets using multi-detector jet images from the CMS Open Data Portal \cite{CMSOpenData}. Jet classification is a well-studied problem in high-energy physics due to its implications for identifying fundamental particle interactions and informing experimental designs at particle accelerators \cite{Andrews_2022}.

This research represents a step forward in quantum-enhanced machine learning, particularly for tasks in high-energy physics. By leveraging QONNs within a transformer architecture, we aim to advance the capabilities of QML for processing high-dimensional datasets efficiently. Our model evaluation using CMS Open Data shows that QONNs offer both efficient computation and strong classification performance, highlighting their potential for practical use in physics and other fields.

\section{Architectures and Circuits}

The architecture of our Quantum Vision Transformer (QViT) leverages a hybrid quantum-classical approach, extending the traditional vision transformer by embedding quantum orthogonal neural networks (QONNs) into its key components. Unlike classical transformers, which rely on entirely classical attention mechanisms and multi-layer perceptrons (MLPs), our model incorporates quantum orthogonal layers to enhance computational efficiency and the overall performance of attention-based operations on high-dimensional data. Inspired by recent advances in quantum transformers, where parametrized quantum circuits are utilized to encode data and perform orthogonal transformations, we adopt an architecture that replaces classical fully connected layers with quantum circuits, thereby allowing us to perform attention mechanisms within the quantum space. This quantum adaptation introduces orthogonal transformations through quantum-specific circuits, such as the pyramid circuits, which facilitate stable and efficient training by preserving gradient properties.

Our architectural approach begins with patch extraction and embedding, as in the classical vision transformer, but diverges in the attention mechanism and where QONNs play a central role. In the following subsections, we detail each stage of the architecture, covering patch extraction, self-attention with quantum orthogonal layer circuits, and MLPs.

\subsection{Vision Transformers}

Vision Transformers (ViTs) have redefined image classification tasks by employing an attention-based mechanism that processes images as a sequence of smaller patches rather than relying on convolutional operations. By utilizing self-attention mechanisms, ViTs efficiently capture both local and global dependencies, making them highly effective for high-dimensional datasets. Our Quantum Vision Transformer (QViT) adapts the ViT architecture by embedding Quantum Orthogonal Neural Networks (QONNs) into its attention mechanism to enhance computational efficiency and performance, particularly for jet image classification.

\begin{figure}[ht]
    \centering
    \includegraphics[width=\textwidth, trim=0 550 0 0, clip]{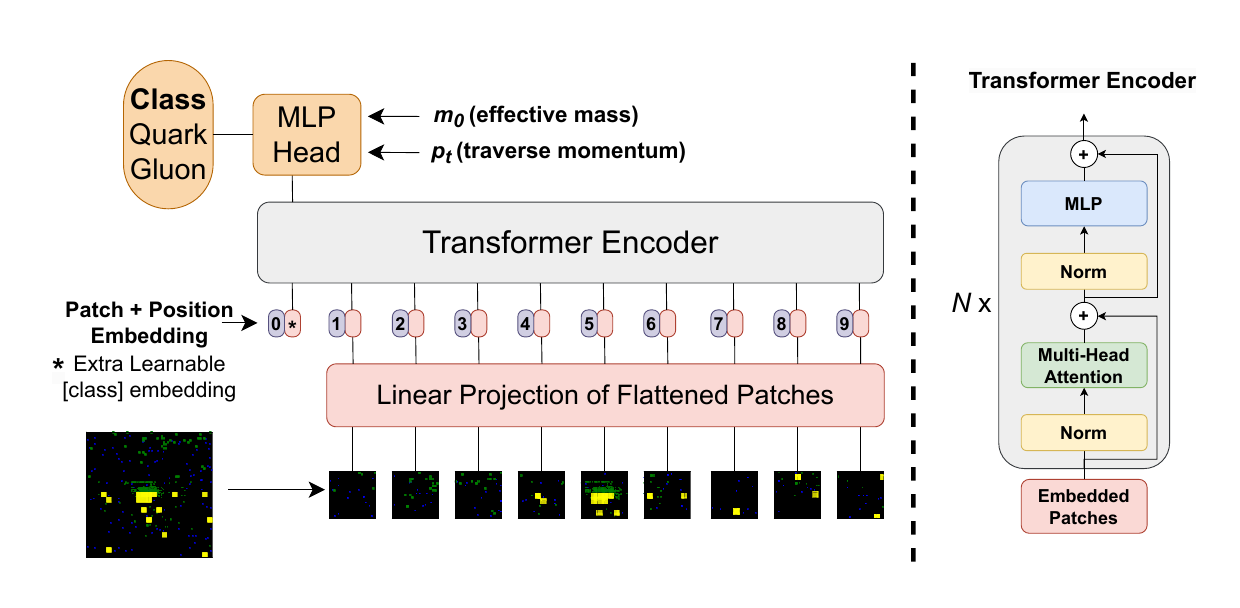}
    \caption{Architecture of the Vision Transformers proposed.}
    \label{fig:vision_transformer}
\end{figure}

\paragraph{Patch Extraction and Embedding}
The input image of size $H \times W \times C$ (height, width, channels) is divided into a grid of $N$ non-overlapping patches, each of size $P \times P$. Each patch is flattened into a 1D vector $\mathbf{x}_i \in \mathbb{R}^{P^2 \cdot C}$, where $i \in \{1, 2, ..., N\}$. These patch vectors are linearly projected into a fixed-dimensional embedding space of size $D$:
\[
\mathbf{z}_i = \mathbf{E} \cdot \text{Flatten}(\mathbf{x}_i),
\]
where $\mathbf{E} \in \mathbb{R}^{D \times (P^2 \cdot C)}$ is a learnable embedding matrix. A learnable class token $\mathbf{z}_{\text{cls}} \in \mathbb{R}^D$ is appended to the sequence of patch embeddings to enable classification. Additionally, positional encodings $\mathbf{p}_i \in \mathbb{R}^D$ are added to retain spatial information:
\[
\mathbf{z}_i' = \mathbf{z}_i + \mathbf{p}_i, \quad i \in \{1, ..., N\}.
\]

\paragraph{Transformer Encoder}
The sequence of embeddings, including the class token, is processed by a stack of $L$ transformer encoder layers. Each layer consists of two sub-layers: Multi-Head Self-Attention (MHSA) and a feed-forward network (FFN). The self-attention mechanism computes attention scores to determine the importance of each patch in the context of others:
\[
\text{Attention}(\mathbf{Q}, \mathbf{K}, \mathbf{V}) = \text{softmax}\left(\frac{\mathbf{Q} \mathbf{K}^\top}{\sqrt{d_k}}\right) \mathbf{V},
\]
where $\mathbf{Q}$, $\mathbf{K}$, and $\mathbf{V}$ are the query, key, and value matrices derived from the input embeddings. Positional encodings ensure that spatial relationships are preserved during this computation.

The output of the transformer encoder is a refined set of patch embeddings and the updated class token $\mathbf{z}_{\text{cls}}$. The class token, which aggregates global information from the input patches, is used for downstream classification.

\paragraph{Classification with Auxiliary Features}
For binary classification, the class token $\mathbf{z}_{\text{cls}}$ is concatenated with two auxiliary jet-level features, the effective mass $m_0$ and transverse momentum $p_T$:
\[
\mathbf{h} = [\mathbf{z}_{\text{cls}}; m_0; p_T].
\]
This combined vector is passed through a Multi-Layer Perceptron (MLP) with a final sigmoid activation to predict the binary class label:
\[
\hat{y} = \sigma(\mathbf{W} \cdot \mathbf{h} + \mathbf{b}),
\]
where $\mathbf{W}$ and $\mathbf{b}$ are the weights and biases of the MLP, and $\sigma(\cdot)$ denotes the sigmoid function. This setup enables effective classification of quark- and gluon-initiated jets while leveraging both image-based and auxiliary features.

\subsection{Quantum Circuits}

In our Quantum Vision Transformer (QViT) architecture, the quantum circuits are specifically designed to implement orthogonal transformations that are integral to the quantum orthogonal neural networks (QONNs) used in the model. These circuits utilize Reconfigurable Beam Splitter (RBS) gates for orthogonal transformations, data loading circuits to prepare quantum states, and quantum layers structured to compute attention coefficients. This section details the various components of the quantum circuits in our model, including the RBS gates, vector loading circuits, orthogonal layer circuits, and the attention coefficient circuit.

\paragraph{RBS Gates}
The RBS gate is a fundamental building block in our quantum circuit, acting as a two-qubit gate with a single tunable parameter, \(\theta\), which controls the rotation in the two-dimensional subspace defined by the basis \(\{|01\rangle, |10\rangle\}\). The RBS gate performs a rotation given by:
\[
RBS(\theta) = 
\begin{pmatrix}
1 & 0 & 0 & 0 \\
0 & \cos \theta & \sin \theta & 0 \\
0 & -\sin \theta & \cos \theta & 0 \\
0 & 0 & 0 & 1 \\
\end{pmatrix}.
\]
This gate swaps the states \(|01\rangle\) and \(|10\rangle\) with amplitudes \(\cos \theta\) and \(\sin \theta\), while acting as the identity on \(|00\rangle\) and \(|11\rangle\). In our circuit design, a network of RBS gates is employed to achieve an orthogonal transformation matrix, forming the basis of the quantum layer in our QViT. Each RBS gate in this network has an independent angle parameter, allowing for flexible control over the transformation. In our design, the RBS gate is implemented through a decomposition involving Hadamard (\(H\)) gates, controlled-Z (\(CZ\)) gates, and single-qubit rotations \(R_y(\pm \theta/2)\). 
%CHHANGE THE IMAGE, YOU NEED CZ GATES AND YOU NEED TO EXPLAIN THE IMAGE IN THE TEXT
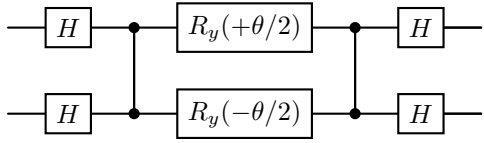
\begin{figure}[h]
    \centering
    \begin{quantikz}
        \lstick{} & \gate{H} & \ctrl{1} & \gate{R_y(+\theta/2)} & \ctrl{1} & \gate{H} & \qw \\
        \lstick{} & \gate{H} & \ctrl{-1} & \gate{R_y(-\theta/2)} & \ctrl{-1} & \gate{H} & \qw
    \end{quantikz}
    \caption{Decomposition of the \(RBS(\theta)\) gate}
    \label{fig:rbs_decomposition}
\end{figure}

\paragraph{Vector Loading Circuits}

To prepare classical data for processing within quantum circuits, we implement vector loading circuits that efficiently map input vectors into quantum states. Specifically, we utilize unary amplitude encoding, where each feature of the input vector \(\vec{x} = (x_0, x_1, \dots, x_{n-1})\) is assigned to a corresponding qubit, ensuring the vector is normalized (\(\|\vec{x}\|_2 = 1\)) for amplitude-based quantum encoding. The vector \(\vec{x}\) is encoded into the following quantum superposition:
\[
|\psi\rangle = x_0|10\cdots 0\rangle + x_1|01\cdots 0\rangle + \cdots + x_{n-1}|0\cdots 01\rangle,
\]
which corresponds to a unary representation where each amplitude is associated with a unique computational basis state \(|i\rangle\).

To achieve this, the circuit begins in the all-zero state \(|\psi_0\rangle = |00\cdots 0\rangle\). The first unary state \(|10\cdots 0\rangle\) is initialized by applying an \(X\) gate to the first qubit. A cascade of \(n-1\) Reconfigurable Beam Splitter (RBS) gates is then applied to progressively entangle the qubits, encoding the amplitudes of \(\vec{x}\) into the quantum state. The rotation angles for the RBS gates, \(\alpha_0, \alpha_1, \dots, \alpha_{n-2}\), are calculated recursively as follows:
\[
\alpha_0 = \arccos(x_0), \quad \alpha_1 = \arccos\left(\frac{x_1}{\sin(\alpha_0)}\right), \quad \alpha_2 = \arccos\left(\frac{x_2}{\sin(\alpha_0) \sin(\alpha_1)}\right), \quad \dots, \quad \alpha_k = \arccos\left(\frac{x_k}{\prod_{j=0}^{k-1} \sin(\alpha_j)}\right),
\]
where \(k \in \{0, 1, \dots, n-2\}\).

\begin{figure}[h]
    \centering
    \begin{quantikz}
        \lstick{\(\ket{0}\)} & \gate{X} & \gate{RBS(\alpha_0)} & \qw                   & \qw                   & \qw \\
        \lstick{\(\ket{0}\)} & \qw      & \ctrl{-1}            & \gate{RBS(\alpha_1)}  & \qw                   & \qw \\
        \lstick{\(\ket{0}\)} & \qw      & \qw                  & \ctrl{-1}             & \gate{RBS(\alpha_2)}  & \qw \\
        \lstick{\(\ket{0}\)} & \qw      & \qw                  & \qw                   & \ctrl{-1}             & \qw
    \end{quantikz}
    \caption{4-Qubit Vector Loading Circuit with Reconfigurable Beam Splitter (RBS) Gates.}
    \label{fig:vector_loader}
\end{figure}
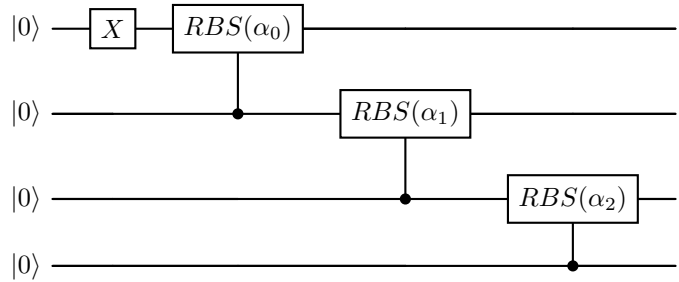

The resulting circuit efficiently prepares the desired quantum state \(|\psi\rangle\) with only \(n-1\) RBS gates. An example for an 4-dimensional input vector is illustrated in Figure~\ref{fig:vector_loader}, where the unary states are sequentially loaded using RBS gates. This structure ensures compatibility with subsequent quantum orthogonal transformations, while maintaining a linear circuit depth.

The normalization of the input vector \(\vec{x}\) ensures compatibility with quantum operations, as the amplitudes must satisfy \(\|\vec{x}\|_2 = 1\). While this constraint may seem restrictive, it does not degrade the model’s performance. Orthogonal transformations, such as those implemented by quantum layers, are norm-preserving, ensuring that the learned representation is unaffected by the normalization. Additionally, normalizing all input vectors to the same norm is analogous to standard preprocessing practices in classical machine learning, where data is scaled to a uniform range.

\paragraph{Orthogonal Layer Circuit}
The orthogonal layer circuit in our model is constructed as a pyramid-like structure of RBS gates, each with tunable parameters that independently control the transformation angles. This pyramidal structure mimics a fully connected neural network layer with orthogonal weights, enabling efficient implementation of orthogonal matrices in quantum space. Each row of RBS gates in the circuit contributes to a progressively complex entanglement pattern, allowing for high-dimensional transformations with fewer parameters than classical equivalents. This orthogonal layer serves as a replacement for the linear projection layers in the classical ViT, bringing quantum computational advantages in both efficiency and resource management.
\begin{figure}[h]
    \centering
    \begin{quantikz}
        \lstick{\(\ket{0}\)} & \gate{RBS(\theta_1)} & \qw                 & \gate{RBS(\theta_3)} & \qw                 & \gate{RBS(\theta_6)} & \qw \\
        \lstick{\(\ket{0}\)} & \ctrl{-1}            & \gate{RBS(\theta_2)} & \ctrl{-1}            & \gate{RBS(\theta_5)} & \ctrl{-1}            & \qw \\
        \lstick{\(\ket{0}\)} & \qw                  & \ctrl{-1}           & \gate{RBS(\theta_4)} & \ctrl{-1}           & \qw                  & \qw \\
        \lstick{\(\ket{0}\)} & \qw                  & \qw                 & \ctrl{-1}            & \qw                 & \qw                  & \qw
    \end{quantikz}
    \caption{Pyramid Circuit with RBS Gates}
    \label{fig:pyramid_circuit}
\end{figure}
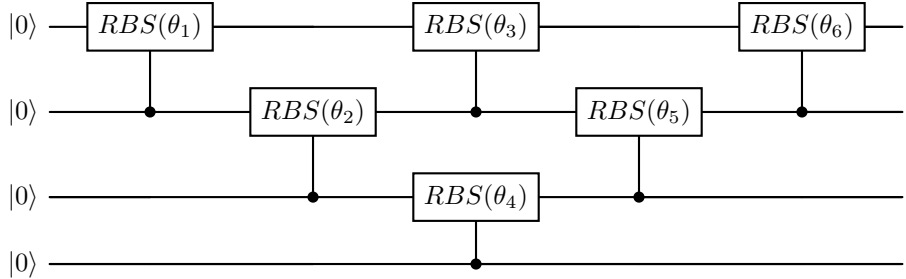

The pyramid structure of the circuit ensures efficient implementation of orthogonal transformations. For an \( n \times n \) orthogonal matrix, the circuit uses \(\frac{n(n-1)}{2}\) RBS gates, matching the degrees of freedom of an orthogonal matrix. Each row of RBS gates adds complexity to the entanglement, enabling high-dimensional transformations with fewer parameters compared to classical implementations.

\paragraph{Attention Coefficient Circuit}
To compute attention coefficients in our QViT model, we employ a quantum circuit that calculates the dot product between the query and key vectors using quantum orthogonal layers. This circuit leverages data loaders for both the input vectors and a single attention coefficient. Each coefficient, \(A_{ij} = \vec{x_i}^T W \vec{x_j}\), is calculated by first loading \(\vec{x_i}\) and \(\vec{x_j}\) into the circuit, followed by the orthogonal layer \(W\). The probability of measuring the output in the state \(|1\rangle\) of the first qubit corresponds to the attention score. 

This approach allows the attention mechanism to be implemented as a quantum operation, with non-linearity introduced through quantum measurement. The resulting coefficients are further processed to form the attention map, enabling the model to selectively focus on relevant features across patches in the input image.
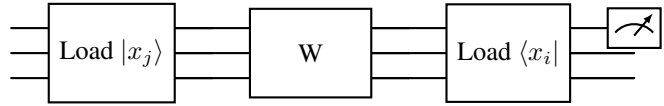
\begin{figure}[H]
    \centering
    \begin{quantikz}[row sep=0.01cm] 
    \setlength{\baselineskip}{1\baselineskip}
        \qw & \gate[3]{\text{Load } |x_j\rangle} & \qw & \gate[3]{\push{\hspace{0.5cm}}\text{W}\push{\hspace{0.5cm}}} & \qw & \gate[3]{\text{Load } \langle x_i|} & \meter{} \\
        \qw &                                     & \qw &             & \qw &                                         & \qw \\
        \qw &                                     & \qw &             & \qw &                                         & \qw \\
    \end{quantikz}
    \caption{Quantum circuit to compute \( |x_i^T W x_j|^2 \), a single attention coefficient, using data loaders for \( x_i \) and \( x_j \), and a quantum orthogonal layer for \( W \).}
    \label{fig:quantum_orthogonal_transformer}
\end{figure}

\section{Experiments}

In this section, we describe the dataset used, the data preprocessing steps, and the hyper-parameters chosen for training our Quantum Vision Transformer (QViT) model. We evaluated our model on a binary classification task of distinguishing between quark-initiated and gluon-initiated jets using multi-detector jet images from the CMS Open Data.

\subsection{Data}

\subsubsection{Data Description}

\begin{figure}[ht]
    \centering
    \includegraphics[width=\textwidth]{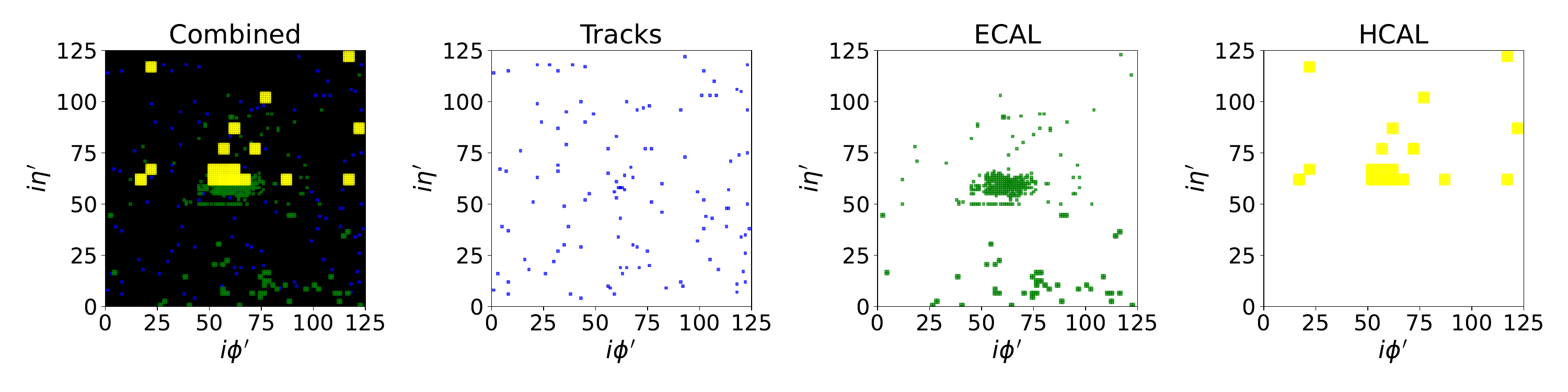}
    \caption{Jet Component Visualization: Tracks, ECAL, HCAL, and Combined Data.}
    \label{fig:jet_image}
\end{figure}

We use the dataset available through the CMS Open Data Portal \cite{CMSOpenData}, which is commonly used in high-energy physics for quark-gluon discrimination tasks. The dataset consists of simulated jet images derived from quantum chromodynamics (QCD) processes, with jets generated and hadronized using the PYTHIA6 Monte Carlo event generator. The dataset includes 933,206 images, each of size $125 \times 125$ pixels, across three channels that represent different components of the CMS detector:
\begin{itemize}
    \item \textbf{Tracks}: Represents the reconstructed trajectories of charged particles passing through the inner tracking system of the CMS detector.
    \item \textbf{Electromagnetic Calorimeter (ECAL)}: Captures energy deposits from electromagnetic interactions, such as those caused by photons and electrons.
    \item \textbf{Hadronic Calorimeter (HCAL)}: Captures energy deposits from hadronic interactions, including those caused by jets of particles containing hadrons.
\end{itemize}

Each image captures the transverse energy depositions within the CMS detector, providing spatial and energy-based information essential for distinguishing between jets. These jets are classified as either quark-initiated or gluon-initiated, with balanced classes in the dataset.

\paragraph{Auxiliary Features: Transverse Momentum (\(p_T\)) and Effective Mass (\(m_0\))}

In addition to the jet image data, two jet-level auxiliary features are included to enhance classification performance:
\begin{itemize}
    \item \textbf{Transverse Momentum (\(p_T\))}: Represents the component of a particle's momentum perpendicular to the beam axis. The \(p_T\) of a jet is a critical feature in particle physics, as it is conserved in high-energy collisions and provides a measure of the jet's energy within the transverse plane.
    \item \textbf{Effective Mass (\(m_0\))}: The effective mass provides an indication of the energy distribution within the jet and its originating particles.
\end{itemize}

Both \(p_T\) and \(m_0\) are essential features for quark-gluon discrimination, as they provide complementary information to the spatial and energy-based patterns within the jet images.

\subsubsection{Data Preprocessing}

To prepare the dataset for input to our QViT model, we perform several preprocessing steps. Initially, we extract a random subset of 50,000 samples for efficient experimentation. The dataset is then split into training (70\%), validation (15\%), and test (15\%) sets.

The pixel intensities of the jet images across the three channels (Tracks, ECAL, HCAL) were already distributed in a well-suited range, requiring no additional normalization or scaling. As such, the image data was used directly without applying min-max scaling to preserve the natural distribution of values.

However, the two auxiliary features—the effective mass ($m_0$) and transverse momentum ($p_T$) of each jet—were normalized using min-max scaling to a range of $[0, 1]$. This scaling was performed independently for the training set, and the same transformation parameters were applied to the validation and test sets to avoid data leakage.

\subsection{Hyper-parameters}

The hyper-parameters for training the QViT model were carefully selected based on a combination of empirical experimentation and established best practices in deep learning for high-energy physics applications. These hyper-parameters were fine-tuned to optimize the model's performance while maintaining computational efficiency. The key hyper-parameters are detailed below:

\begin{itemize}
    \item \textbf{Projection Dimension}: Each patch embedding has a dimension of 8, as larger circuit simulations would have significantly increased training time due to the exponential growth in computational cost.
    \item \textbf{Number of Patches}: Each $125 \times 125$ image is divided into non-overlapping patches of size $25 \times 25$, resulting in 25 patches per image. While this setup preserves spatial information, it also represents a considerable computational challenge, as the simulation requires $26 \times 26$ attention circuits per self-attention block, which already pushes the computational limits of current quantum simulators.
    \item \textbf{Transformer Encoder Blocks}: A single transformer encoder block is used to perform feature extraction.
    \item \textbf{Attention Heads per Block}: A single attention head per block is used, ensuring a simple yet effective self-attention mechanism capable of capturing dependencies across patches without overwhelming computational resources.
    \item \textbf{Dropout Rate}: 0.5, applied to mitigate overfitting by randomly dropping units during training.
\end{itemize}

The model was trained for 15 epochs using the Adam optimizer with a learning rate of $0.0005$ and a batch size of 32, employing binary cross-entropy as the loss function due to the binary nature of the classification task. Training metrics included accuracy and the area under the receiver operating characteristic curve (ROC AUC). This setup allowed for a comprehensive evaluation of the QViT model’s capability to distinguish between quark-initiated and gluon-initiated jets.

\section{Results}

This section compares the performance of the Quantum Vision Transformer (QViT) and the classical Vision Transformer (ViT) on quark-gluon jet classification. The primary goal is to evaluate the effectiveness of quantum attention mechanisms relative to classical ones by analyzing validation AUC, loss across epochs, and test set metrics.

\subsection{Validation Performance}

\begin{figure}[H]
    \centering
    \begin{subfigure}[b]{0.48\textwidth}
        \centering
        \includegraphics[width=\textwidth]{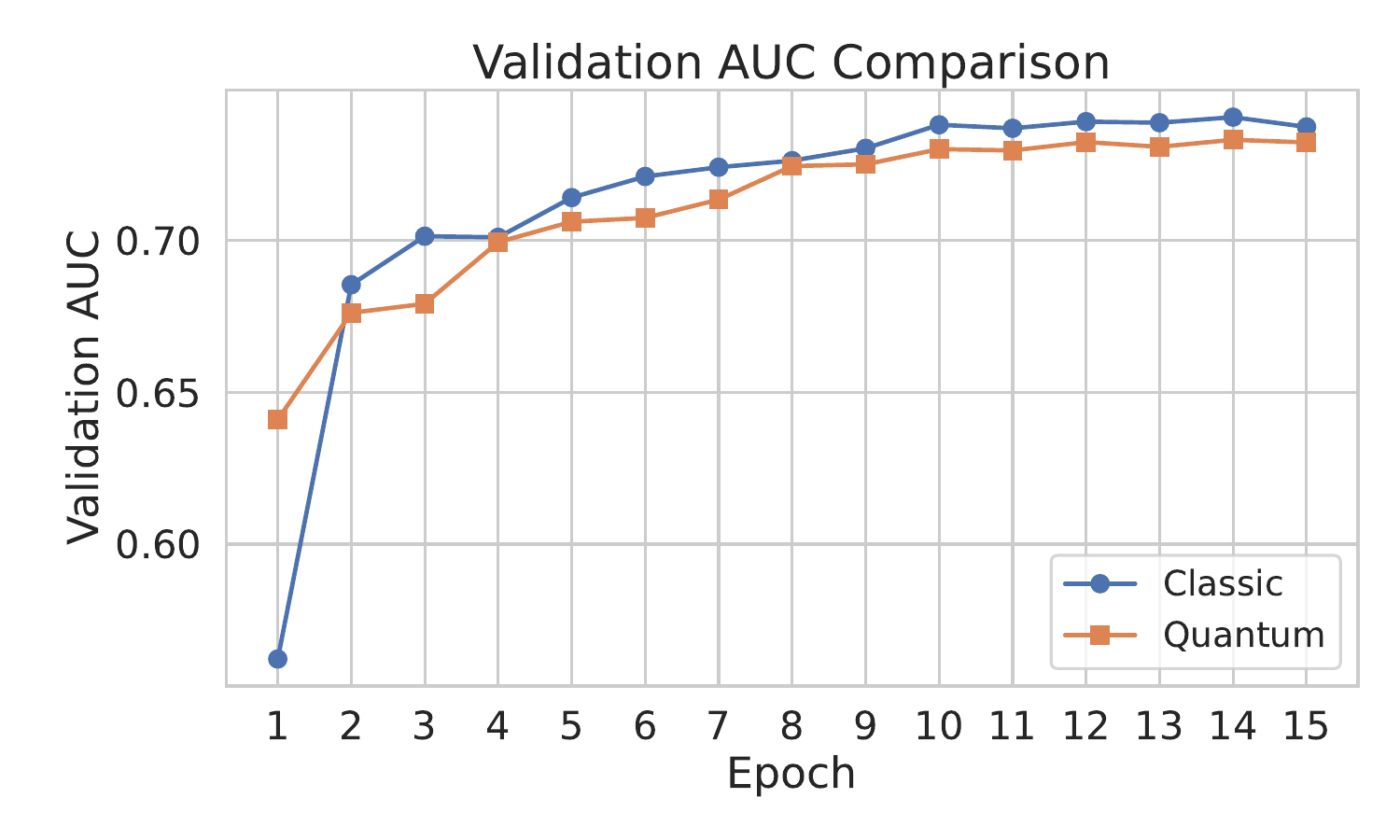}
        \caption{Validation AUC comparison between QViT and classical ViT models.}
        \label{fig:val_auc}
    \end{subfigure}
    \hfill
    \begin{subfigure}[b]{0.48\textwidth}
        \centering
        \includegraphics[width=\textwidth]{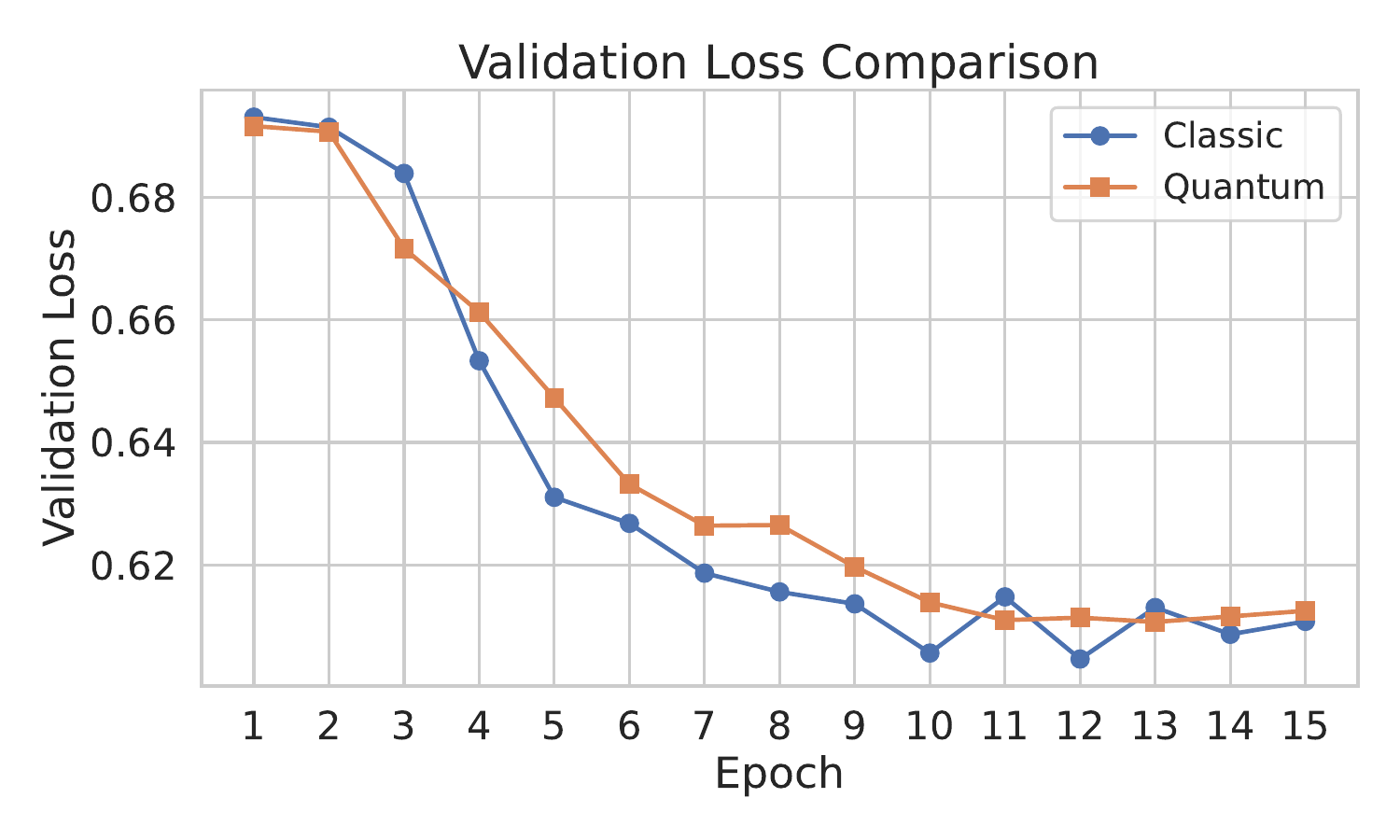}
        \caption{Validation loss comparison between QViT and classical ViT models.}
        \label{fig:val_loss}
    \end{subfigure}
    \caption{Validation metrics for QViT and classical ViT across 15 epochs.}
    \label{fig:val_metrics}
\end{figure}

The validation AUC and loss for the QViT and classical ViT models across 15 epochs are shown in Figure~\ref{fig:val_metrics}. Both models achieve similar validation AUC, converging to approximately 0.675 by the final epoch. This indicates that the quantum attention mechanism in the QViT is as effective as the classical mechanism in distinguishing quark- and gluon-initiated jets.

\subsection{Test Performance}

The final test set performance of the QViT and classical ViT models is summarized in Table~\ref{tab:test_results}. Both models show comparable results, with QViT achieving slightly lower accuracy and AUC but demonstrating similar overall performance. These results highlight the potential of quantum attention mechanisms as a scalable alternative to classical approaches.

\begin{table}[H]
\centering
\caption{Test set performance comparison.}
\label{tab:test_results}
\begin{tabular}{|c|c|c|c|}
\hline
\textbf{Model}       & \textbf{Test Loss} & \textbf{Test Accuracy} & \textbf{Test AUC} \\ \hline
Classical ViT        & 0.6087             & 0.6788                 & 0.7385            \\ \hline
Quantum QViT         & 0.6105             & 0.6755                 & 0.7369            \\ \hline
\end{tabular}
\end{table}

\section{Conclusions}

In this work, we introduced a Quantum Vision Transformer (QViT) model that integrates quantum orthogonal neural networks (QONNs) into the attention mechanism for the challenging task of quark-gluon jet classification in high-energy physics. By embedding quantum circuits into the attention layers, the QViT efficiently processes high-dimensional data while maintaining comparable performance to classical Vision Transformers.

Our analysis demonstrates that quantum attention mechanisms are as effective as classical ones. The QViT achieves validation and test AUC values similar to those of the classical ViT, with minimal differences in accuracy and loss. These results underscore the potential of quantum models for applications in data-intensive fields like high-energy physics, where scalability and efficiency are critical.

Future work will explore enhanced quantum circuit designs and evaluate QViT on larger datasets with more complex tasks. As quantum hardware advances, QViTs could become a practical alternative for machine learning applications, leveraging the unique properties of quantum computing to offer competitive performance with reduced computational resources.

\section{Aknowledgments}
We are thankful to Marçal Comajoan Cara, Cosmos Dong, Roy Forestano, Jogi Suda Neto and Eyup Unlu for useful discussions. 
This research used resources of the National Energy Research Scientific Computing Center, a DOE Office of Science User Facility supported by the Office of Science of the U.S. Department of Energy under Contract No. DE-AC02-05CH11231 using NERSC award NERSC DDR-ERCAP0025759. SG is supported in part by the U.S. Department of Energy (DOE) under Award No. DE-SC0012447. KM is supported in part by the U.S. DOE award number DE-SC0022148. KK is supported in part by US DOE DE-SC0024407. Alessandro Tessi was a participant in the 2024 Google Summer of Code.

\bibliographystyle{unsrt}  
\bibliography{references} 

\end{document}